# POSSIBLE FRACTIONAL QUANTUM HALL EFFECT IN GRAPHITE

Y. Kopelevich[1], B. Raquet[2,3], M. Goiran[2,3], W. Escoffier[2,3], R. R. da Silva[1], J. C. Medina Pantoja[1], I. A. Luk´yanchuk[4], A. Sinchenko[5,6], and P. Monceau[6]

[1]Instituto de Física "Gleb Wataghin", Universidade Estadual de Campinas, UNICAMP 13083-970, Campinas, São Paulo, Brasil

[2]CNRS; UPR 3228 LNCMI; 143 Avenue de Rangueil, F-31400 Toulouse, France

[3]Université de Toulouse, UPS, INSA, LNCMI; F-31400 Toulouse, France

[4]Laboratoire de Physique de Matière Condensée, Université de Picardie Jules Verne, Amiens, 80039, France

[5]Moscow Engineering-Physics Institute, 115409, Moscow, Russia

[6]Institut Néel, CNRS , B.P. 166, 38042 Grenoble cedex 9, France

ABSTRACT

Measurements of basal plane longitudinal $\rho_b(B)$ and Hall $\rho_H(B)$ resistivities were performed on highly oriented pyrolytic graphite (HOPG) samples in pulsed magnetic field up to B = 50 T applied perpendicular to graphene planes, and temperatures 1.5 K ≤ T ≤ 4.2 K. At B > 30 T and for all studied samples, we observed a sign change in $\rho_H(B)$ from electron- to hole-like. For our best quality sample, the measurements revealed the enhancement in $\rho_b(B)$ for B > 34 T (T = 1.8 K), presumably associated with the field-driven charge density wave or Wigner crystallization transition. Besides, well defined plateaus in $\rho_H(B)$ were detected in the ultra-quantum limit revealing the signatures of fractional quantum Hall effect in graphite.

It has been known for a long time that magnetic field B > 20 T applied along the hexagonal c-axis induces in graphite an anomalous high-resistance state (HRS) that can be detected using either basal-plane $\rho_b(B,T)$ or out-of-plane $\rho_c(B,T)$ resistivity measurements [1-9] [20 T >> $B_{QL}$ = 7 - 8 T(QL stands for quantum limit) that pulls all carriers into the lowest Landau level (LLL)]. The boundaries that trace the HRS domain on the B-T plane [5, 7] are in a qualitative agreement with theoretical expectations [10] for the Landau-level-quantization-induced normal metal - charge density wave (CDW) as well as the reentrant CDW-normal metal transitions. However, while the CDW is predicted to occur in the direction of the magnetic field [10], the experimental results [2, 4] indicate the in-plane character of CDW, or formation of two-dimensional (2D) Wigner crystal (WC) state(s) [11, 12]. Supporting either CDW- or WC-based scenarios, the non-Ohmic electrical transport was measured in HRS [4, 8]. Typically, for T = 2 K the HRS emerges in the field interval 25 T < B < 52 T, and the HRS does not occur for T > 10 K [7].

All the HRS studies [1-9] were performed on artificially grown Kish or natural single crystalline graphite samples. To the best of our knowledge, no measurements above 28 T [1] were performed for HOPG, and no Hall resistivity $\rho_H(B)$ measurements above 30 T [6] were made on any type of graphite.

Recent magnetoresistance [13, 14] and scanning tunneling spectroscopy (STS) [15] experiments revealed the integer quantum Hall effect (IQHE) in graphite. The IQHE takes place only in strongly anisotropic (quasi-2D) HOPG samples with the room temperature out-of-plane/basal-plane resistivity ratio $\rho_c/\rho_b > 10^4$, and mosaicity ≤ 0.5° (FWHM obtained from x-ray rocking curves). Together with the high electron mobility μ ~ $10^6$ cm$^2$/Vs [16], this makes HOPG a promising system for the fractional QHE (FQHE) occurrence.

Motivated by these observations, in the present work we studied magnetoresistance in HOPG in pulsed magnetic field up to B = 50 T and 1.5 K ≤ T ≤ 4.2 K. The measurements were performed in Ohmic regime with 300 ms for the total pulse length in AC configuration, at LNCMPI (Toulouse, France). Additional measurements were made using Janis 9T-magnet He-4 cryostat.

Commercially available HOPG samples ZYA and SPI-3 were measured. The sample parameters are: FWHM = 0.4°, $\rho_c/\rho_b$ = 4·10$^4$ ($\rho_b$ = 5 μΩcm and $\rho_c$ = 0.2 Ωcm) for ZYA, and FWHM = 3.5°, $\rho_c/\rho_b$ = 3.8·10$^3$ ($\rho_b$ = 40 μΩcm and $\rho_c$ = 0.15 Ωcm) for SPI-3 HOPG samples (the resistivity data were obtained for B = 0 and T = 300 K). X-ray diffraction (Θ-2Θ) measurements revealed a characteristic hexagonal graphite structure in the Bernal (ABAB…) stacking configuration, with no signature for the rhombohedral phase. The obtained crystal lattice parameters a = 2.48 Å and c = 6.71 Å.

Here we report the results obtained on the ZYA sample of dimensions l x w x t = 2.5 x 2.5 x 0.5 mm$^3$. The magnetic field was applied parallel to the hexagonal c-axis (B ∥ c ∥ t), and $\rho_b(B)$, $\rho_H(B)$ were recorded using the van der Pauw method, sweeping the field between - 50 T and + 50 T.

From the data presented in Fig. 1 (a, b), one observes that $\rho_b(B)$ goes through the maximum at $B_{m1}$ = 18 T, develops two local minima at $B_\alpha$ = 30 T and $B_{\beta1}$ = 34 T, and passes through the second maximum at $B_{m2}$ ≈ 43 T. Thus, $\rho_b(B)$ represents all characteristic features reported for Kish graphite [7], where e. g. the resistivity minima at $B_\alpha$ = 28 T and $B_{\beta1}$ = 33 T, attributed to multiple field-induced CDW phases, were measured at T = 1.7 K. The onset of HRS in Kish graphite is accompanied by a rapid decrease of $\rho_H(B) \sim \sigma_H(B)$ = -e($n_e$ – $n_h$)/B [6], where $n_e$ and $n_h$ are majority electron and hole carrier densities, respectively. At low enough temperatures, $\rho_H(B)$ tends to zero as B approaches ~ 30 T, suggesting that $\rho_H(B)$ may change its sign from "minus" to "plus" with a further field increasing [6]. Our results (Fig. 1, a) give the experimental proof that the sign of $\rho_H(B)$ changes at $B_H = 43$ T. The straightforward explanation of this effect would be the carrier density imbalance change from $n_e > n_h$ ($B < B_H$) to $n_h > n_e$ ($B > B_H$). This provides us with a new insight on the resistivity drop taking place at $B > B_{m2}$. Namely, one assumes that decrease of both $\rho_b(B)$ and $-\rho_H(B)$ at $B > B_{m1}$ originates from the hole density increase [1], whereas the HRS is due to the field-induced Wigner crystallization of electrons. Then, nonmonotonic $\rho_b(B)$ can be simply understood using the equation for parallel resistors $\rho_b = \rho_{be}\rho_{bh}/(\rho_{be}+\rho_{bh})$, [$\rho_{be}(B)$ and $\rho_{bh}(B)$ are electron and hole basal-plane resistivities, respectively], without invoking any reentrant transition in the electronic state (noting, $\rho_b \gg \rho_H$).

We also measured the similar sign reversal in $\rho_H(B)$ at B ~ 30 - 35 T for two SPI-3 samples. However, due to a poorer quality of those samples, neither negative magnetoresistance nor HRS were detected. Instead, $\rho_b(B)$ saturates for B > 18 T.

Next, we focus our attention on plateau-like and oscillatory features in $\rho_H(B)$ and $\rho_b(B)$, seen in Figs. 1-4 as a fine structure.

In Fig. 2 we plotted $\Delta\rho_b(B)$ vs. 1/B for B < 5 T, where $\Delta\rho_b(B)$ is obtained after subtraction of the monotonic background resistivity $\rho_b^{bg}(B)$: the data clearly demonstrate that the fine structure is due to Shubnikov-de-Haas (SdH) oscillations. The obtained period of SdH oscillations $\Delta(B^{-1}) = 0.208 \pm 0.004$ T$^{-1}$ (the frequency $B_0 = 4.8 \pm 0.1$ T) corresponds to the extremal cross section of the Fermi surface of the majority electrons [17].

The analysis of experimental results obtained for $B < B_{QL}$ [13, 18], showed that electrons mainly contribute to the measured $\rho_H(B)$, whereas the contribution from Dirac-like majority holes is tiny [18]. Thus, the measured IQHE staircase is consistent with either conventional massive electrons with Berry´s phase 0, or chiral massive electrons having Berry´s phase $2\pi$, as in graphene bilayer [19, 20]. We stress that IQHE staircases measured for HOPG [13, 18] and graphene bilayer samples [20] overlap when plotted as a function of the filling factor $\nu = B_0/B$ [18], testifying on the quasi-2D nature of HOPG. The inset in Fig. 2 illustrates the QH plateau occurrence at $\nu = 1$ and $\nu = 2$. In the same figure we show $\rho_H(\nu)$ measured [13] for HOPG-UC (Union Carbide Co.) sample. It can be readily seen that the main IQHE plateau is centered at B = $B_0$ for both ZYA and UC HOPG samples, demonstrating the universality in the behavior of these strongly anisotropic samples. The Hall resistivity $\rho_H(\nu = 1) = 0.82$ m$\Omega$cm for ZYA HOPG was obtained from the measured Hall resistance $R_H(\nu = 1) = 16.5$ m$\Omega$, assuming the uniform current distribution through the sample thickness t = 0.5 mm. Taking the distance between neighboring graphene planes d = c/2 = 3.355 Å, one gets N $\approx 1.5 \cdot 10^6$ independent graphene (bi)layers contributing to the measured signal. This gives the Hall resistance per (bi)layer $R_H (\nu = 1) = N \cdot R_H(\nu = 1) \approx 24.8$ k$\Omega$, that practically coincides with the Klitzing fundamental Hall resistance $h/e^2 \cong 25.8$ k$\Omega$. However, because of the strong sample anisotropy ($\rho_c/\rho_b = 4 \cdot 10^4$), the measuring current can be concentrated within the effective sample thickness $t_{eff} < t$ [21], implying that the actual value of $R_H (\nu = 1)$ can be smaller. Taking the QH plateau sequence $R_H = h/4\nu e^2$ as predicted (and measured) for graphene bilayer [19, 20], and the measured difference $\Delta\rho_H(\nu) = \rho_H(\nu = 1) - \rho_H(\nu = 2) \approx 0.2$ m$\Omega$cm (Fig. 2, inset), one estimates the effective thickness of the electron "layers" $l_{eff} \cong 6.2$ Å, responsible for IQHE. Interestingly, the obtained value of $l_{eff}$ agrees very well with the c-axis lattice parameter c = 6.71 Å, resembling the theoretical result for IQHE in bulk graphite [22]. Whether this is an accidental coincidence or it has a deeper reason, remains to be seen.

The data presented in Fig. 3 demonstrate that plateaus in $\rho_H(B)$ take also place for $\nu \ll 1$. As Fig. 3 (a, b, c) exemplifies, plateaus are centered quite accurately at $\nu$ = 2/7, 1/4, 2/9, 1/5, 2/11, 1/6, 2/15, 1/8, 2/17, 1/9. It appears, that all these numbers correspond to the filling factors $\nu$ = 2/m (m = 1, 2, 3,…) proposed by Halperin [23] for the case of bound electron pairs, i. e. 2e-charge bosons. In principle, the existence of 2e-bosons in the ultraquantum limit can be justified assuming the electron pairing driven by the Landau level quantization [24]. $\Delta\rho_H(B)$ steps between neighboring plateaus agree with the FQHE scenario, as well. For instance, $\Delta\rho_H(B) \approx$ 0.22 m$\Omega$cm measured between $\nu$ = 1/4 and $\nu$ = 2/7 (Fig. 3 a) plateaus, coincides with the expected value $\Delta\rho_H(B)$ = (h/8e$^2$)·c$\approx$ 0.216 m$\Omega$cm (here we assume unbroken spin and valley degeneracies in graphite).

It is worth to note, that $l_{eff} \approx$ c = 6.71 Å is much smaller than the magnetic length $l_B$[Å] = $(\hbar/eB)^{1/2}$ = 250/B$^{1/2}$[T$^{1/2}$] in the whole studied field range. Thus, recent 3D models for FQHE [25, 26], probably relevant to bulk bismuth [27, 28], do not apply to highly anisotropic graphite.

On the other hand, one may argue against the QHE in both bulk graphite and bismuth, because $\rho_b(B)$ does not vanish in the plateau region, and $\rho_b(B) > \rho_H(B)$. However, small dips and not vanishing of the longitudinal resistivity $\rho_{xx}(B) > \rho_{xy}(B)$ were measured, e. g., for Bechgaard salt (TMTSF)$_2$PF$_6$ [29], Bi$_{2-x}$Sn$_x$Te$_3$ and Sb$_{2-x}$Sn$_x$Te$_3$[30], η-Mo$_4$O$_{11}$ [31] layered crystals, as well as for GaAs/AlGaAs 2DES [32, 33], in both IQHE [29-31] and FQHE [32, 33] regimes. In particular, in Ref. [33] FQH states resulting from the melted Wigner crystal were detected at very high global longitudinal resistance level of R$_{xx} \sim$ 1 M$\Omega$.

Figure 4 illustrates the correlation between QH plateaus and dips in $\Delta\rho_b(B)$ measured in the present work. Fig. 4 also demonstrates that minima in $\Delta\rho_b(B)$ are somewhat shifted from the plateau centers which is the characteristic feature of QHE in bulk materials [30, 31]. Thus, it is legitimate to treat the data obtained on graphite in a similar way. For $\nu > 2/7$, no correlation between dips in $\Delta\rho_b(B)$ and plateau-like features in $\rho_H(B)$ is found, and no plateaus corresponding to Halperin´s $\nu = 2/m$ fractional filling factors can be unambiguously identified. Further studies should clarify this observation.

In summary, we report the results of basal-plane Hall resistivity $\rho_H(B)$ and longitudinal resistivity $\rho_b(B)$ measurements performed on HOPG samples up to B = 50 T. The sign change in the Hall resistivity from electron- to hole-type in ultraquantum limit is reported for graphite for the first time. For our best quality samples, FQHE associated with majority electrons is detected for filling factors $\nu << 1$, and ascribed to a quantum liquid of 2e-bosons [23]. The obtained results provide evidence that strongly anisotropic graphite can be considered as a system of quasi-2D layers of the thickness $l_{eff} \approx c = 6.71$ Å that exhibit independent integer ($\nu \geq 1$) or fractional ($\nu < 1$) quantum Hall states.

We thank K. Behnia, D. V. Khveshchenko, A. H. MacDonald, and C. Morais Smith for useful discussions. This work was partially supported by FAPESP, CNPq, CAPES, COFECUB, FP7-IRSES-ROBOCON, INTAS grant N° 05-1000008-7972, ANR grant BLAN07-03-192276 and the LIA between RAS and CNRS.

**FIGURE CAPTIONS**

Fig. 1. (Color online) (a) Basal-plane $\rho_b(B)$ and Hall $\rho_H(B)$ resistivities measured at T = 1.8 K; $\rho_H(B)$ changes sign at $B_H$ = 43 T; (b) $\rho_b(B)$ demonstrates nonmonotonic behavior discussed in the text.

Fig. 2. (Color online) Shubnikov de Haas resistivity oscillations with the period $\Delta(B^{-1})$ = 0.208 ± 0.004 $T^{-1}$ corresponding to the majority electrons; $\Delta\rho_b$ is obtained subtracting the monotonic background resistivity $\rho_b{}^{bg}(B)$. The inset demonstrates quantum Hall plateaus measured for HOPG-ZYA and HOPG-UC [$\rho_H(B)/3.6$] samples; $\nu = B_0/B$ ($B_0$ = 4.68 T for HOPG-UC, and $B_0$ = 4.8 ± 0.1 T for HOPG-ZYA).

Fig. 3. (Color online) Quantum Hall plateaus observed for various fractional filling factors $\nu = B_0/B$ ($B_0$ = 4.8 ± 0.1 T).

Fig.4. (Color online) Plateaus in the Hall resistivity $\rho_H(B)$ correlate with the minima in $\Delta\rho_b$ (multiplied by factor 5 and arbitrary shifted along the vertical axis); dotted lines mark centers of QH plateaus.

Fig. 1

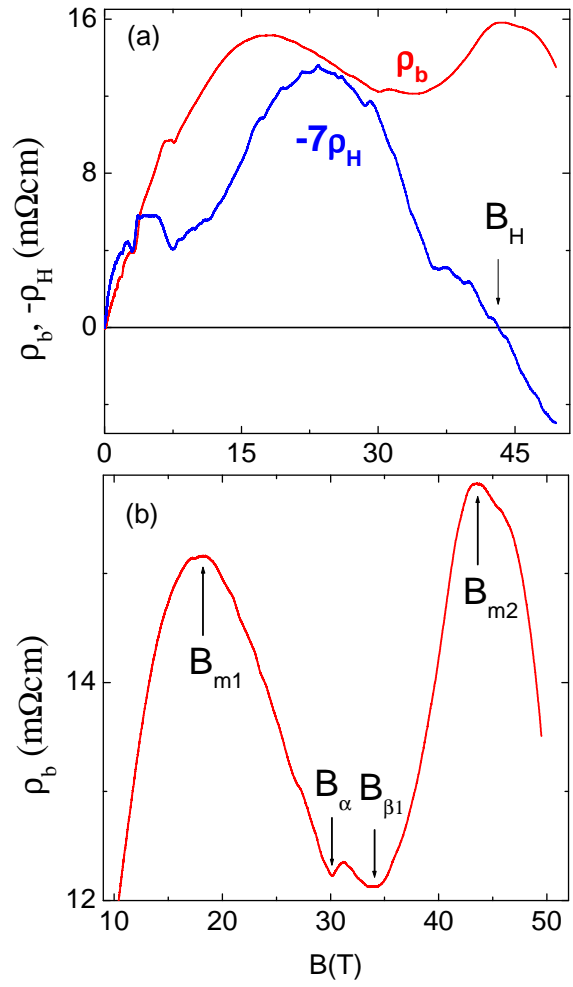

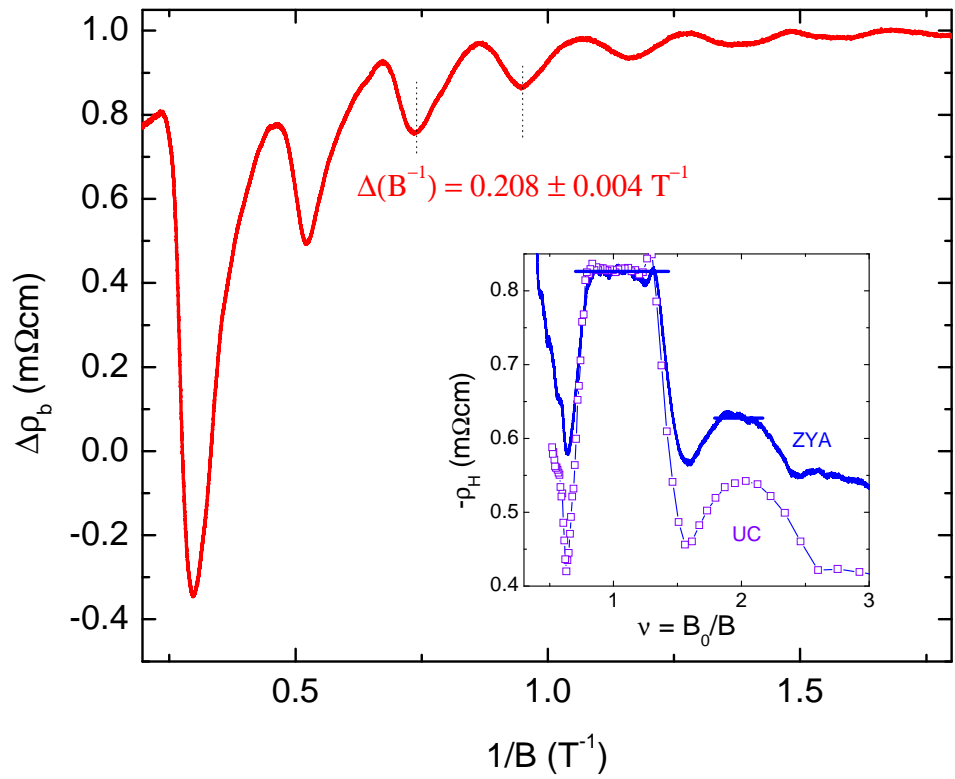

$\Delta(B^{-1}) = 0.208 \pm 0.004 \text{ T}^{-1}$

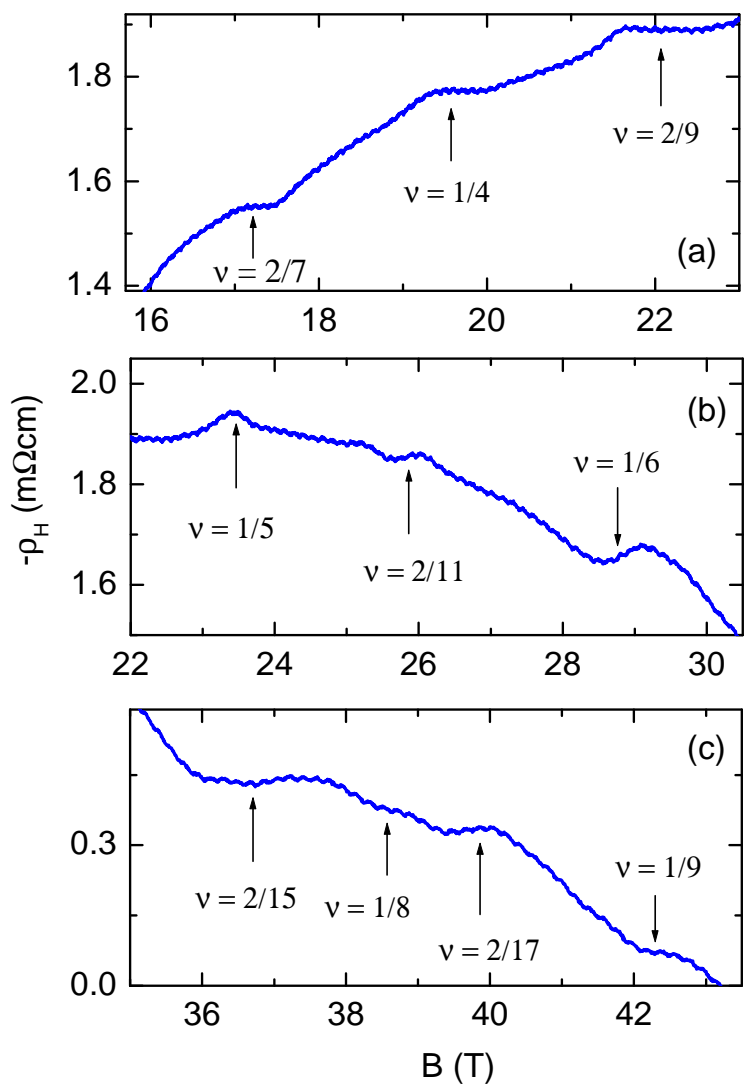

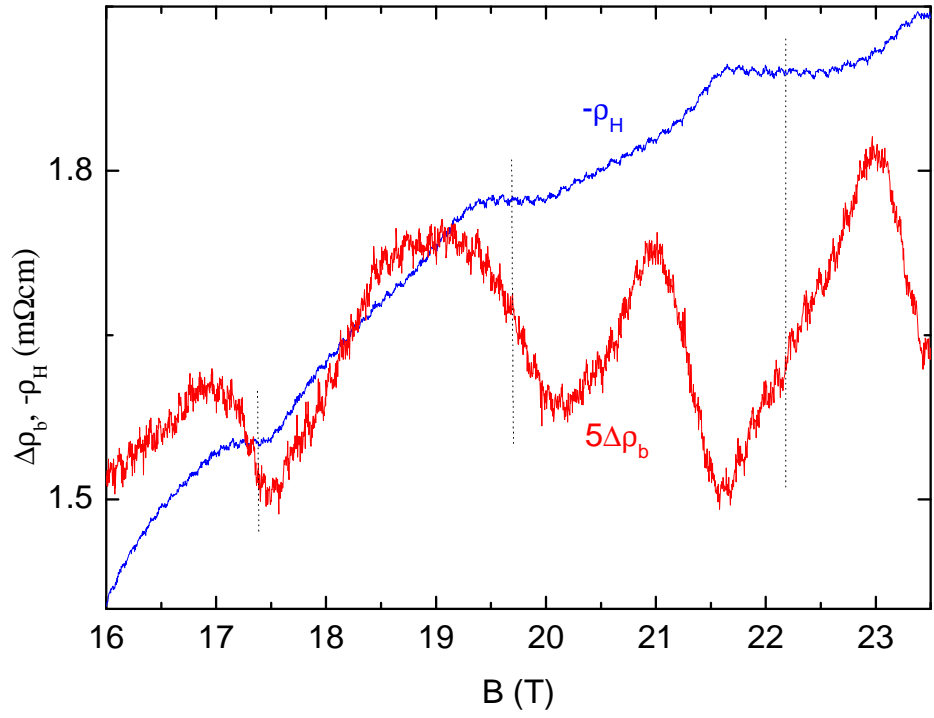